# Simultaneous Segmentation and Relaxometry for MRI through Multitask Learning

Peng Cao[1], Jing liu[1], Shuyu Tang[1], Andrew Leynes[1], Janine M. Lupo[1], Duan Xu[1], Peder E. Z. Larson[1]

[1]Department of Radiology and Biomedical Imaging, University of California at San Francisco, San Francisco, CA, USA

*Correspondence to:

Peng Cao

Department of Radiology and Biomedical Imaging, University of California at San Francisco, San Francisco, CA, USA

Address: 1700 4th Street, San Francisco CA 94158

Phone: 415-341-3762

Email: peng.cao@ucsf.edu

**Abstract**

**Purpose:** This study demonstrated an MR signal multitask learning method for 3D simultaneous segmentation and relaxometry of human brain tissues.

**Materials and Methods:** A 3D inversion-prepared balanced steady-state free precession sequence was used for acquiring in vivo multi-contrast brain images. The deep neural network contained 3 residual blocks, and each block had 8 fully connected layers with sigmoid activation, layer norm, and 256 neurons in each layer. Online synthesized MR signal evolutions and labels were used to train the neural network batch-by-batch. Empirically defined ranges of T1 and T2 values for the normal gray matter, white matter and cerebrospinal fluid (CSF) were used as the prior knowledge. MRI brain experiments were performed on 3 healthy volunteers. The mean and standard deviation for the T1 and T2 values in vivo were reported and compared to literature values. Additional animal (N=6) and prostate patient (N=1) experiments were performed to compare the estimated T1 and T2 values with those from gold standard methods and to demonstrate clinical applications of the proposed method.

**Results:** In animal validation experiment, the differences/errors (mean difference ± standard deviation of difference) between the T1 and T2 values estimated from the proposed method and the ground truth were 113 ± 486 and 154 ± 512 ms for T1, and 5 ± 33 and 7 ± 41 ms for T2, respectively. In healthy volunteer experiments (N=3), whole brain segmentation and relaxometry were finished within ~5 seconds. The estimated apparent T1 and T2 maps were in accordance with known brain anatomy, and not affected by coil sensitivity variation. Gray matter, white matter, and CSF were successfully segmented. The deep neural network can also generate synthetic T1 and T2 weighted images.

**Conclusion:** The proposed multitask learning method can directly generate brain apparent T1 and T2 maps, as well as synthetic T1 and T2 weighted images, in conjunction with segmentation of gray matter, white matter and CSF.

**Introduction**

Creating an MR signal multitask learning paradigm [1] is an important step towards building an intelligent MRI machine that could better interpret complex MR data. Segmentation and relaxometry are two important tasks in MRI. The MRI relaxometry allows direct measurements of physical parameters in normal and pathological brain tissues in vivo [2–4] with applications in tissue segmentation [5–7]. Recent MRI relaxometry studies showed that segmentation based on absolute parameter values is intrinsically highly tolerable to the intensity variations in MR images, and is less dependent on the bias field correction step which is essential in the conventional intensity-based segmentation methods [8]. Besides, the combination of multi-parameter MRI relaxometry techniques, such as T1/T2 relaxometry, can provide more robust tissue classification and comprehensive characterization of brain morphology. The clinical applications of the joint segmentation and relaxometry include the automatic detection and quantification for T1/T2 abnormalities such as lesion, and volumetric changes such as atrophy.

Rapid acquisition methods for MRI relaxometry based on the balanced steady-state free precession sequence (bSSFP) or its modifications have been intensively investigated [9–12]. A previous study showed the feasibility of combining a rapid variable flip angle gradient echo scan and a bSSFP scan for simultaneous T1 and T2 mapping, i.e., DESPOT1 and DESPOT2 [11]. In that study, the T1 values estimated from the variable flip angle gradient echo images were used to solve the signal equation of bSSFP for estimating T2 values [11]. Some other studies jointly estimated T1 and T2 values for the inversion-prepared bSSFP (IR-bSSFP) sequence, modeling the IR-bSSFP signal evolution as a simplified exponential function [9, 10]. It is also possible to estimate T1 and T2 values from bSSFP signal evolutions by dictionary matching or deep neural network [12–14]. A recent MR fingerprinting (MRF) study, based on a variable-parameter/heterogeneous bSSFP acquisition, showed the possibility of using a dictionary to solve the Bloch equation for deriving T1 and T2 values which are assumed to align with grid entries in the parameter space [12]. Studies also showed the feasibility of using the neural network for MRF quantification [13, 14]. Recent advances of quantitative imaging technique encourage a paradigm shift from acquiring intensity-based images to a direct measurement of tissue MR parameters; meanwhile, they also present an apparent complexity in the data analysis.

In biological tissues, T1 and T2 values are often positively associated with each other, e.g., the T1 and T2 values of normal gray matter are both longer than those of normal white matter [15, 16], reflecting intrinsic properties of the tissue composition and molecular environment. Therefore, it is feasible and of interest to develop a multitask learning paradigm [1] that is based on the known MR signal evolution model, i.e., the solution of Bloch equation, and the association of MR parameters with tissue types. Specifically, as illustrated in Figure 1, by training the neural network with a large amount of multi-parameter MRI signal evolutions and tissue type labels, the deep learning algorithm can "learn" the tissue-signal-parameter association. Furthermore, using parallel computing on a graphic processing unit (GPU), hundreds of MRI signal evolutions could be simulated in seconds for each training batch, offering a virtually unlimited number of samples for training the neural network. In summary, the deep learning technique offers a novel tool for exploiting the shared information in the segmentation and relaxometry based on the modeling of complex MRI signal evolutions and tissue-parameter associations in biological tissues.

This study demonstrated an MR signal multitask learning paradigm for 3D simultaneous segmentation and relaxometry of human brain tissues through a deep neural network. Ranges of T1 and T2 values for the normal gray matter, white matter and cerebrospinal fluid (CSF) were used as the prior knowledge. The proposed method can directly generate brain apparent T1 and T2 maps in conjunction with segmentation of gray matter, white matter, and CSF. Additional animal and prostate patient experiments were performed to compare the estimated T1 and T2 values with those from gold standard methods and to demonstrate clinical applications of the proposed method.

**Methods**

<u>Neural network modeling of tissue type probabilities</u>

In the present study, we used a multilayer perceptron (MLP) [17, 18] with a parameter sharing output [1] for the joint regression (i.e., T1, T2, and B0 estimation) and classification (i.e., segmentation) task. We extended MLP to a residual network structure [19] to improve its capacity in modeling while avoiding the "vanishing" of gradient during the training of neural network [19]. We followed the conventional definition of probabilities of gray matter, white matter and CSF in one voxel [20]. Assume the observed MRI signal evolution for $m$ inversion times at voxel $i$ is $y_i = (y_{i1}, y_{i2}, \ldots y_{im})$. Then, the tissue classes form a mixel of random variables $x_i = \{x_{i1}, x_{i2}, \ldots x_{iN}\}$ that satisfies $\sum_{n=1}^{N} x_{in} = 1$ and $x_{in} \geq 0$ for $n = 1$ to $N$ (and $N$ for the number of tissue classes). The probability distribution function for the observation $y_i$ at voxel $i$, conditioned on the true mixel $x_i$, $P(y_i|x_i)$, can be approximated by the neural network. We further assumed a uniform distribution of the T1 and T2 parameters within the feasible region to simulate $x_i$ and $y_i$ (as shown in Fig. 1), as those parameters for different tissues could only be within specific ranges [15, 16]. In addition, in this study, we chose the apparent T1 and T2 ranges empirically based on the initial results from dictionary matching on in vivo data, and empirically optimized the T1 and T2 ranges in the experiment. We did such manual optimization to find the appropriate ranges for T1 and T2, i.e., the best separation of the T1 and T2 values for different tissue types. To summarize, in the present study, the probabilities of gray matter, white matter, and CSF were calculated as the probabilities of measured MRI signal evolutions from the tissue types, i.e., fall in ranges of parameters for specific tissue types [15, 16].

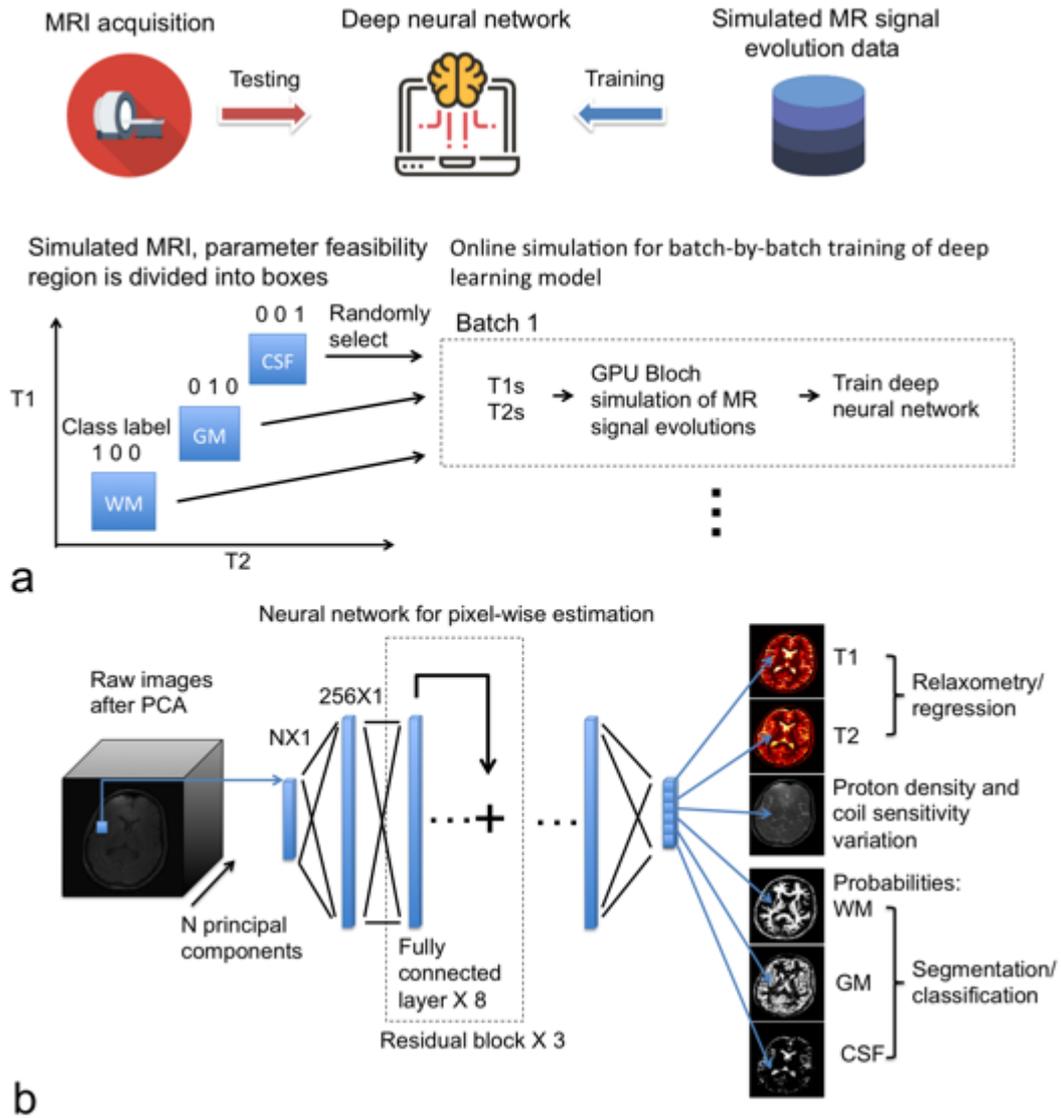

**Figure 1 (a**) MRI signal synthetic model in the brain is used to train the neural network; while in vivo MRI data is used to test the neural network. Specifically, the proposed method assumed the feasible parameter space could be divided into boxes that were binary labeled for white matter (WM), gray matter (GM), and cerebrospinal fluid (CSF). **(b)** A deep neural network was used to perform pixel-wise estimation of T1, T2, B0 (not shown) and proton density/M0 (weighted by coil sensitivity variation) as a regression output, and WM, GM and CSF probabilities as a classification output. Each input pixel contained coefficients of principal components from a principal component analysis (PCA) preprocessing. The neural network contains 3 residual blocks, and each block has 8 fully connected layers, i.e., an MLP with "by-pass" connections to avoid gradient vanishing during training. The neural network was trained with T1 and T2 values from the parameter feasibility region as well as varied proton density weightings, varying noise levels, and binary class labels (in Fig. a).

Neural network design, training, and testing

The deep neural network contained 3 residual blocks (in Fig. 1b). Each residual block contains 8 fully connected layers (in Fig. 1b) with sigmoid activation, one layer norm and 256 neurons in each layer. In the final layer, different activation functions were used for the regression and classification outputs, i.e., sigmoid and softmax, respectively. The deep neural network was implemented in the TensorFlow software package (https://www.tensorflow.org/). We also provide source code in the "MRIPY" toolbox (https://github.com/larsonlab/mripy).

Online synthesized MR signal evolutions and labels were used to train the neural network batch-by-batch (in Fig. 1a). Within each batch, uniformly randomized T1 and T2 values, frequency offsets (i.e., B0 inhomogeneity), proton densities, binary-labels of tissue/fluid type and varying noise levels were used to simulate MR signal evolutions at 3T. The T1 and T2 values were uniformly distributed within the ranges: 450 ms < T1 < 825 ms and 25 ms < T2 < 62.5 ms for white matter, 750 ms < T1 < 1125 ms and 55 ms < T2 < 92.5 ms for gray matter, and 2250 ms < T1 < 3750 ms and 125 ms < T2 < 275 ms for CSF, respectively. The T1 and T2 ranges were chosen empirically and were fixed for all experiments in this study. In addition, the proton density, noise level, and frequency/B0 offset were uniformly distributed within the ranges: 0 to 100%, 0 to 0.2% (relative to the equilibrium magnetization) and 0 to $1/(2 \times TR)$, respectively. Furthermore, these "inversion recovery" signal evolution curves were compressed/pre-processed by applying principal component analysis (PCA) to reduce the dimensionality of data. We computed the PCA basis of the theoretical temporal evolution curves, i.e., the dictionary used in MRF approach, followed by the projection of the data to the basis. The PCA reduces the inversion time dimension to principal components of 6 for the first volunteer or 20 for the other volunteers. The cost function for training neural network was designed as a mean squared difference between the output of the neural network and the known parameters and class labels, and ADAM optimizer was used [21]. Training and testing were performed on an NVIDIA TITAN X GPU. Other details of training are 600 to 800 samples per batch, ~2 seconds per batch simulation/training speed, and >18000 batches for a typical 10 hours of training. The training time was 5 to 10 hours and the testing time was ~5 seconds (for a typical image matrix of $160 \times 160 \times 64$ and 20 principal components).

MRI sequence and in vivo data acquisition

Animal validation experiments were performed on a Bruker 3T MRI (Bruker BioSpin GmbH, Germany). Six Sprague Dawley rats (N = 6) were scanned at postnatal day 30. Within them, two rats were controls, and the other four rats had local ischemic infarction in the right hemisphere of the brain, which was induced at postnatal day 10. For conventional T1 mapping, an inversion prepared fast spin echo (IR-FSE) sequence was used with parameters: 8 inversion times from 7 to 2100 ms, TE/TR = 80/3500 ms, RARE factor = 32, matrix size = 128 × 128, and resolution = 0.23 × 0.23 $mm^2$. For conventional T2 mapping, a Carr-Purcell Meiboom-Gill (CPMG) sequence was applied with parameters: 16 TEs from 16 to 256 ms, TR = 1730 ms, matrix size = 128 × 128, and resolution = 0.23 × 0.23 $mm^2$. Parameters for IR-bSSFP included FA = 30° and 60°, TE/TR = 2/4 ms, 128 × 128, and resolution = 0.23 × 0.23 $mm^2$, and one frequency offset for FA = 30° and no frequency offset for FA = 60°. The two-flip-angle scheme enabled the estimation of the B1 map, accounting for the B1 variation in experiments.

Human experiments were approved by the Institutional Review Board. All experiments were performed on a GE Excite 3-T clinical MRI system (GE Healthcare, Waukesha, WI) with 40 mT/m strength, 150 mT/m/ms slew rate gradients. An inversion-prepared balanced steady-state free precession (IR-bSSFP) sequence [9, 10] in combination with parallel imaging and compressed sensing reconstruction was used for acquiring the dynamic images in vivo [22, 23]. Three healthy volunteers and one patient volunteer with prostate cancer have been imaged using the sequence. For the first healthy volunteer, parameters for IR-bSSFP included flip angle (FA) = 30°, echo time/repetition time (TE/TR) = 1.5/4.3 ms, FOV = 28 cm, image matrix = 196 × 192 × 30, resolution = 1.4 × 1.1 × 2.0 $mm^3$ and no frequency offsets [22, 23]. For the other two healthy volunteers, the FOV and image matrix were slightly different from the first one, with FOV = 22 cm, image matrix = 160 × 160 × 64, resolution = 1.4 × 1.4 × 2.6 $mm^3$, and 2 scans were acquired with no frequency offset and offset of 1/(2×TR) respectively [22, 23]. For the patient volunteer with prostate cancer, we used the same IR-bSSFP imaging protocol as was used in the last healthy volunteer experiment. In the IR-bSSFP sequence, after each inversion pulse, data acquisition was set to be 3 s using CIRCUS undersampling strategy [22, 23], which allowed pseudo-random variable-density sampling with a spiral-like trajectory and golden-ratio profile on the Cartesian ky-kz plane. Images were reconstructed at 13 different inversion times for the first volunteer, and 30 inversion times for the other two volunteers. The acceleration factor was moderate (R=1.3 to 1.6) to first prove the concept of the experiment. The parallel imaging and compressed sensing reconstruction were performed with the use of L1-ESPIRIT method on the BART software (v3.01, https://mrirecon.github.io/bart/). All the processing for neural network was done in the magnitude image data.

Additonally, conventional T1 mapping and T2 mapping were also acquired on the third volunteer. For T1 mapping, a spoiled gradient echo (SPGR) sequence with two FAs, 6 and 12 degrees was applied [24], with TE/TR = 1.5/5.2 ms, matrix size = 160 × 160 × 94, and resolution = 0.8 × 0.8 × 2.5 mm$^2$. For T2 mapping, the CPMG sequence was used with parameters: 8 TEs ranging from 6.25ms to 50 ms, TR = 1500 ms, matrix size = 160 × 160, 11 slices, resolution = 0.8 × 0.8 mm$^2$, and slice thickness = 3 mm. In addition, a T1 weighted anatomical image was acquired using a standard magnetization prepared rapid gradient echo (MPRAGE) protocol with parameters: TE/TR = 1.9/5.5 ms, inversion time = 450 ms, matrix size = 512 × 512 × 76, and resolution = 0.4 × 0.4 × 2.5 mm$^3$.

Signal evolution of IR-bSSFP at the transient phase and simplifications in simulation

The signal evolution of IR-bSSFP sequence has been well described in previous studies [9, 10]. We chose to use IR-bSSFP because of its simplicity in implementation, and its enhanced contrast by mixing the T1 and T2 relaxations in the signal evolution. We used GPU-based parallel computing to simulate bSSFP signal evolution in real-time for training the neural network, as shown in Figure 1. In the simulation algorithm, we used simplifications for the excitation and procession in one TR, and they are given as:

Excitation: $R_{ex}(\alpha, \phi) = R_x(\phi) R_z(\alpha) R_x(-\phi)$ and $M_{k,+} = R_{ex}(\alpha, \phi)M_{k,-}$      Eq. 1a

where $R_x$ and $R_z$ are the standard SO(3) rotation matrices, $\alpha$ denotes the flip angle, and $\phi$ is the phase of RF, $M_{k,-}$ and $M_{k,+}$ is the magnetizations before or after $k$th excitation respectively, and

Procession: $M_{k+1,-} = diag([E_2, E_2, E_1])R_z(\varphi)M_{k,+} + m_0[0, 0, 1 - E_1]$      Eq. 1b

where $E_{1,2} = e^{-TR/T1,2}$, $m_0$ proton density, $\varphi$ phase accumulation due to the frequency offset during procession period and $M_{k,+}$/$M_{k+1,-}$ the magnetization at the beginning or ending time points of the k$^{th}$ TR. The TR-by-TR evolution was sequentially computed, while different spins were simulated in parallel on GPU with a batch size of 100. On typical GPU hardware, e.g., NVIDIA TITAN X or 1070/1060 GTX, simulating hundreds of signal evolutions with 600 to 1000 TRs normally could be done within a few seconds.

Data analysis

For comparison, in the first experiment, a 3D Markov random field method using Gaussian mixture model (modified from a 2D "image segmentation based on Markov random fields" code at https://www.mathworks.com/matlabcentral/fileexchange/33592-image-segmentation-based-on-markov-random-fields?requestedDomain=true) [25] was applied to T1 and T2 maps from the result of the neural network. The open source code was for 2D image segmentation, we changed the 2D operation to 3D operation, while, still using the same inference method. Markov random field and Gaussian mixture models have been widely used as the major framework in image segmentation methods [20, 25, 26]. In the present study, the Markov random field was initialized by k-means clustering, and no spatial smoothing or bias field correction was applied (since the input T1 and T2 maps from neural network output contained no significant intensity variations). The single-task learning was tested based on the current neural network, but without the segmentation output, i.e., the neural network was used to estimate the T1 and T2 maps with or without limitations in the parameter space. In the second experiment, dictionary-element like quantization, which is similar to the dictionary match approach [12], was performed on the same in vivo data with step sizes of 500 ms, 50 ms, and 7 Hz for T1, T2 and B0, respectively. In addition, the skull stripping and standard brain segmentation were also performed using BET and FAST algorithms in FSL software [27–29]. In the last experiment, the BET and FAST algorithms were applied to the high-resolution 3D magnetization prepared rapid gradient echo (MPRAGE) image to achieve optimal segmentation performance. We also simulated MPRAGE, spin echo and fluid-attenuated inversion recovery (FLAIR) images from the T1, T2 and proton density/M0 (with coil sensitivity variations removed after the bias field correction) acquired by the IR-bSSFP scan and the proposed neural network relaxometry. On the raw proton density map, the 3D bias field correction was performed using our in-house developed method as reported in [30]. In this study, the tissue segmentation was fixed in the bias field estimation, which is different from the original method described in [30].

**Results**

Figure 2 shows the comparison of the proposed method with T1 and T2 mapping by standard parameter mapping methods (i.e., IR-FSE and CPMG) on six rats (N = 6, all images and analysis are presented in Supporting Figure 1) scanned at postnatal day 30. There is a strong spatial agreement between the methods. In Figures 3 to 6, the neural network quantified T1 and T2 maps showed a reduction of quantification artifacts and improvement in the separation of gray and white matter. Figure 3 shows that the estimated apparent T1 and T2 maps were in accordance with known brain anatomy. Varying coil sensitivity, as well as modest artifacts due to the aliasing and Gibbs ringing along slice dimension (shown in raw images after PCA in Fig 3 top left), were decoupled from the apparent T1 and T2 maps and segmentation results. In addition, the deep brain gray matter nuclei, e.g., caudate nucleus, were well recognized by the deep neural network. For segmentation comparison, a 3D Markov random field was applied to T1 and T2 maps from the neural network. The classification result from Markov random field showed some ambiguities in detecting detailed white matter structures (Fig. 3, arrow) that were not present in the neural network direct segmentation. In this dataset, an offset frequency scan was not acquired, and the B0 map was not estimated; however, no major bSSFP banding artifacts were noticed over the 6 cm slab coverage of this scan. The typical result from bias field correction on the proton density map was shown in Supporting Figure 2.

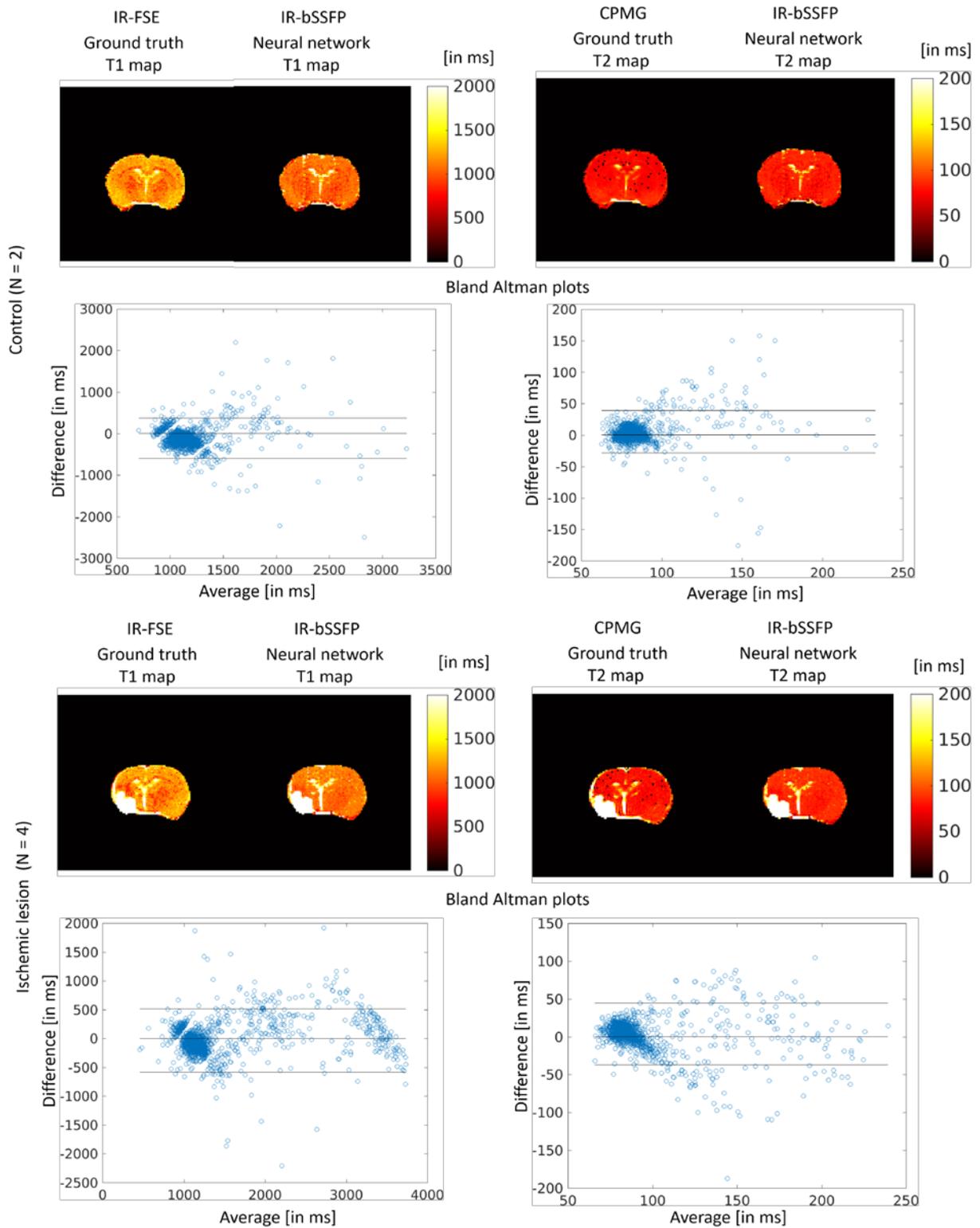

**Figure 2** The comparison of the proposed method with T1 and T2 mapping by standard parameter mapping methods (i.e., IR-FSE and CPMG) on six Sprague Dawley rats (N = 6 with two normal controls and four with chronic ischemic lesions in the left hemisphere, all images and analysis are presented in Supporting Figure 1). There is a strong spatial agreement between the methods. The differences/errors (mean difference ± standard deviation of difference) in two Bland Altman plots were 113 ± 486 and 154 ± 512 ms for T1,

and 5 ± 33 and 7 ± 41 ms for T2, respectively. The on average shorter T1 and T2 values resulting from IR-bSSFP with neural network processing can be attributed to magnetization transfer effects [31–33].

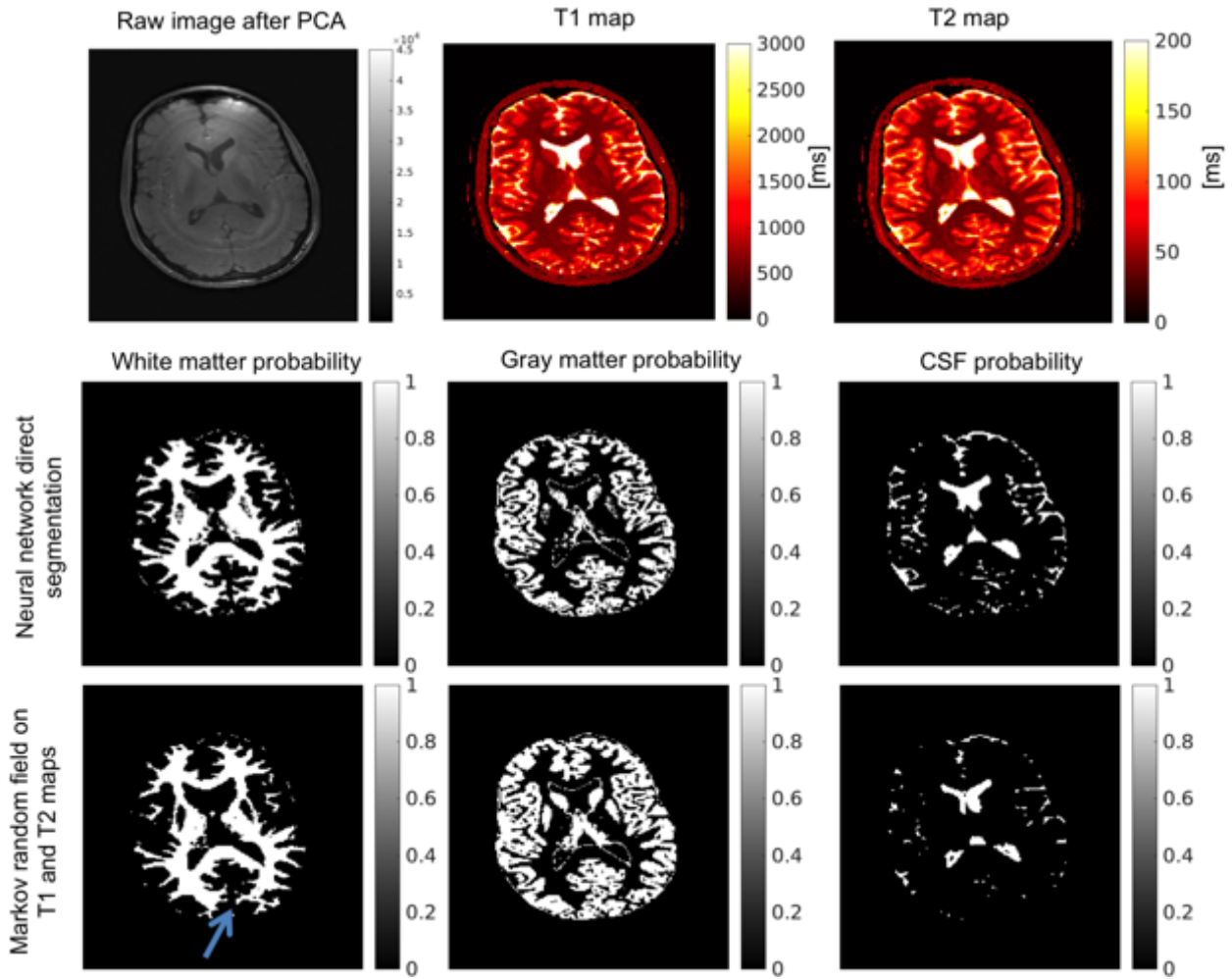

**Figure 3** The segmentation and relaxometry results for one slice (out of 30 slices). The data were acquired with 3D IR-bSSFP, 13 inversion times, and no frequency offsets. Notice that the proposed method tolerated varied coil sensitivity profiles and the modest artifacts in the raw images caused by the aliasing and Gibbs ringing along slice dimension (top left). The estimated T1 and T2 maps ~~agreed well with brain anatomy~~ are consistent with our knowledge of brain anatomical structure in MRI. The 3D Markov random field algorithm was applied to T1 and T2 maps for a segmentation comparison. The classification result from Markov random field showed some ambiguities in detecting detailed white matter structures (arrow), likely due to the error propagation from T1 and T2 maps. The skull stripping was performed using the BET algorithm in FSL.

Figure 4 shows the comparison of the quantification results from multi-task deep neural network with T1 and T2 range limitations, single-task deep neural network with T1 and T2 range limitations, dictionary-element-like quantization on T1, T2, and B0 (similar to the MR fingerprinting approach), as well as the naïve single-task approach (without T1 and T2 limitations). In this study, two IR-bSSFP scans with varied frequency offsets were acquired, which allowed for the estimation of a B0 map in both neural network and dictionary matching methods. The single-task and dictionary-element like quantization were vulnerable to artifacts in apparent T1, T2, and B0 maps, especially in the high B0 variation areas. On the other hand, the deep neural network can produce relatively accurate apparent T1 and T2 maps as well as continuously valued B0 maps.

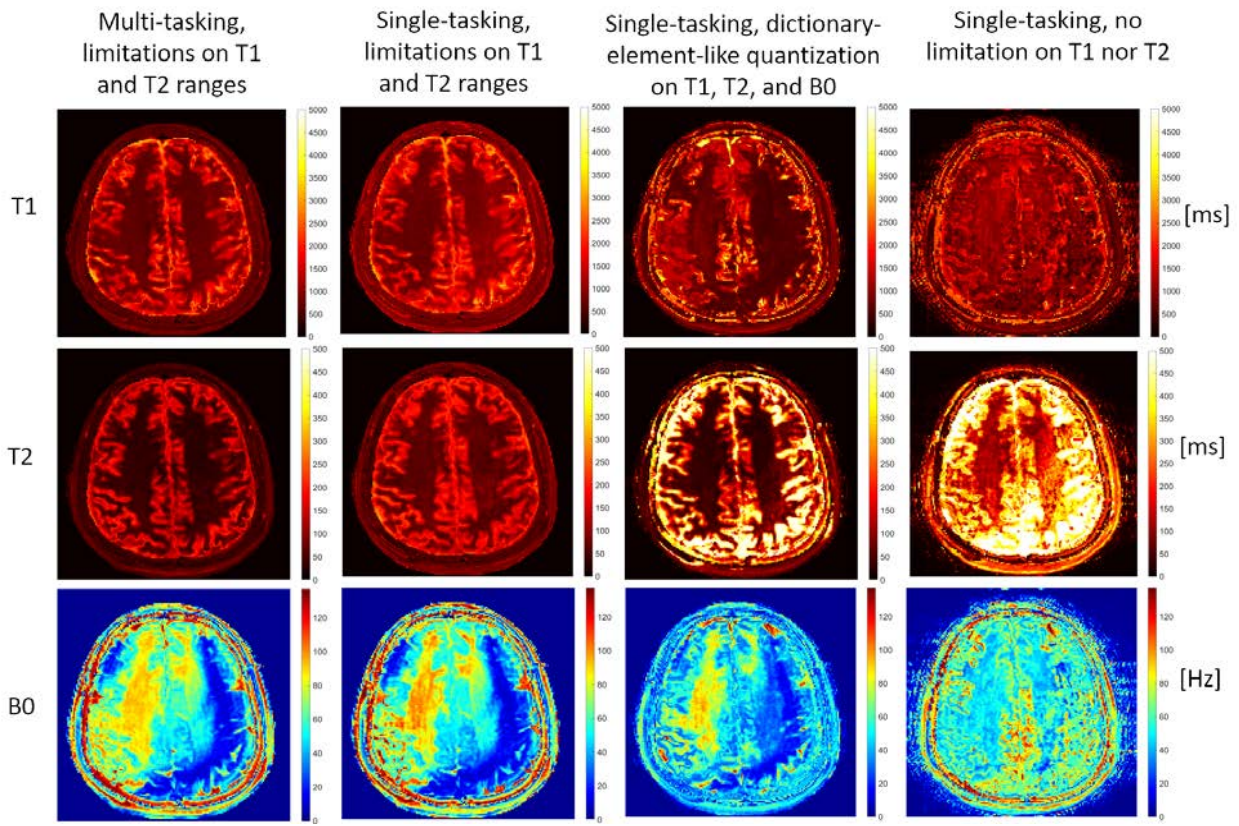

**Figure 4** In vivo apparent T1, T2 and B0 estimation using the proposed multi-task deep neural network with T1 and T2 range limitations, single-task deep neural network with T1 and T2 range limitations, and dictionary-element-like quantization on T1, T2, and B0, as well as the naive single-task approach with different levels of T1 and T2 limitations. Data were acquired with 3D IR-bSSFP, 30 inversion times, and 2

different frequency offsets at 0 and 1/(2×TR). The single-task and dictionary-element like quantization were vulnerable to artifacts in T1, T2, and B0 maps. On the other hand, the deep neural network outputs with T1 and T2 range limitations appeared more accurate relative to brain anatomy and provided more continuous T1, T2 and B0 values.

Figure 5 shows a comparison of the proposed method with standard T1 and T2 mapping results. In Figure 5, the apparent T1 and T2 maps from IR-bSSFP with the neural network were qualitatively similar to the standard ones, although with shorter T1 and T2 values on average. This was likely caused by magnetization transfer effects from the inversion pulse and bSSFP readout [31–33].

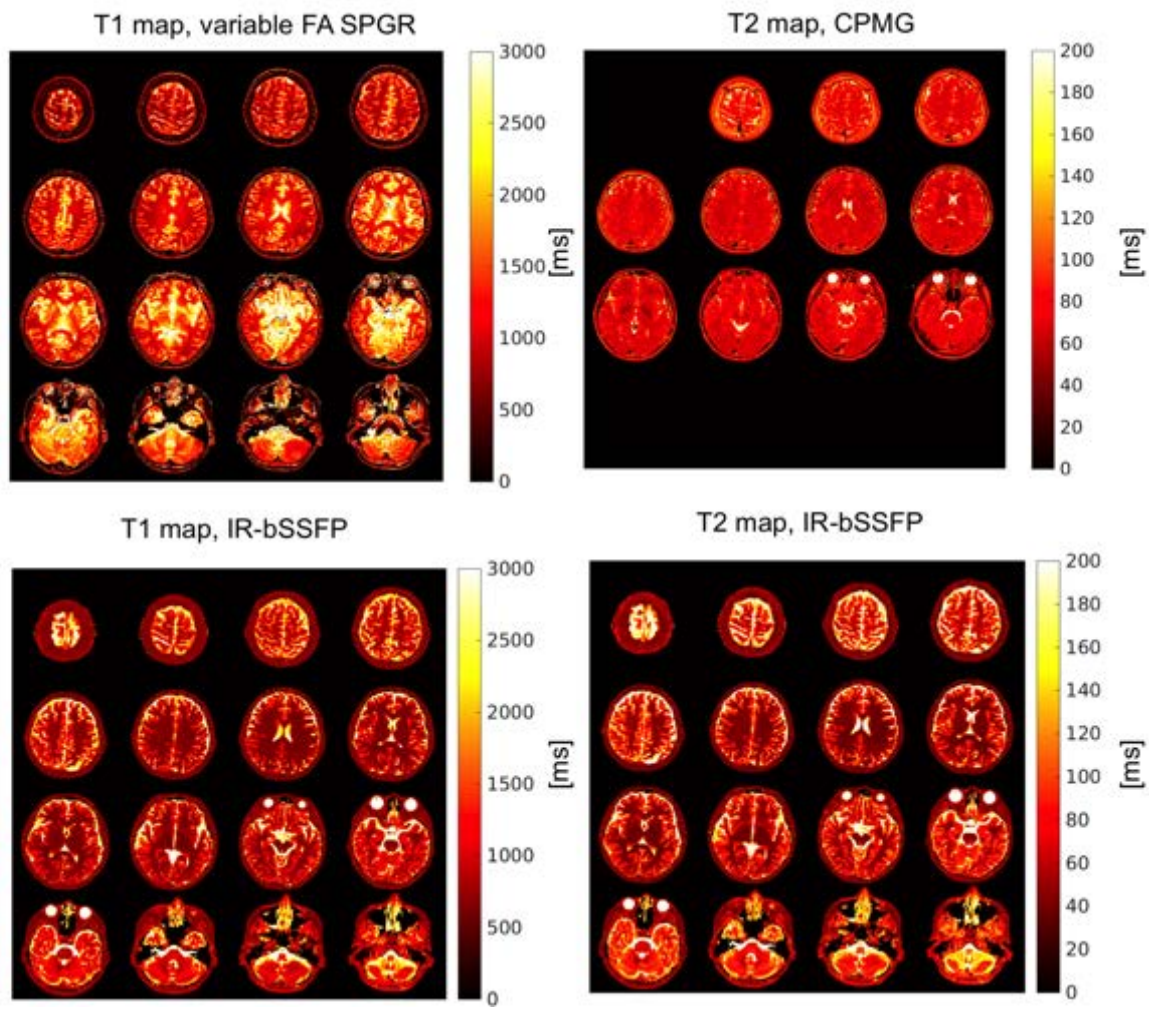

**Figure 5** Comparison of the proposed method with T1 and T2 mapping by standard parameter mapping methods (top row). There is a strong structural agreement between the methods. The on average shorter T1 and T2 values resulting from IR-bSSFP with neural network processing can be attributed to magnetization transfer effects [31–33].

In Figure 6, the neural network segmentation matched the FSL FAST segmentation results in general, with slightly improved gray matter separation in the cortex (arrows in Fig. 6); while FAST is better at some deep brain structures, i.e., putamen and the head of caudate. In Figure 7, preliminary results show that the typical three MRI contrast weighted images, MPRAGE, spin echo, and FLAIR were simulated based on the neural network results. The simulation showed the potential of performing multi-contrast 3D MRI scans (including T1, T2, and FLAIR) using one image sequence. In addition, the apparent T1 and T2 values estimated were summarized in Table 1. The average T1 and T2 values in three tissue types were relatively consistent across three healthy volunteers. The smaller T1 and T2 values compared with those from standard T1 and T2 mapping (in Fig. 5), and literature values [15, 16] were likely caused by the magnetization transfer effect [31–33].

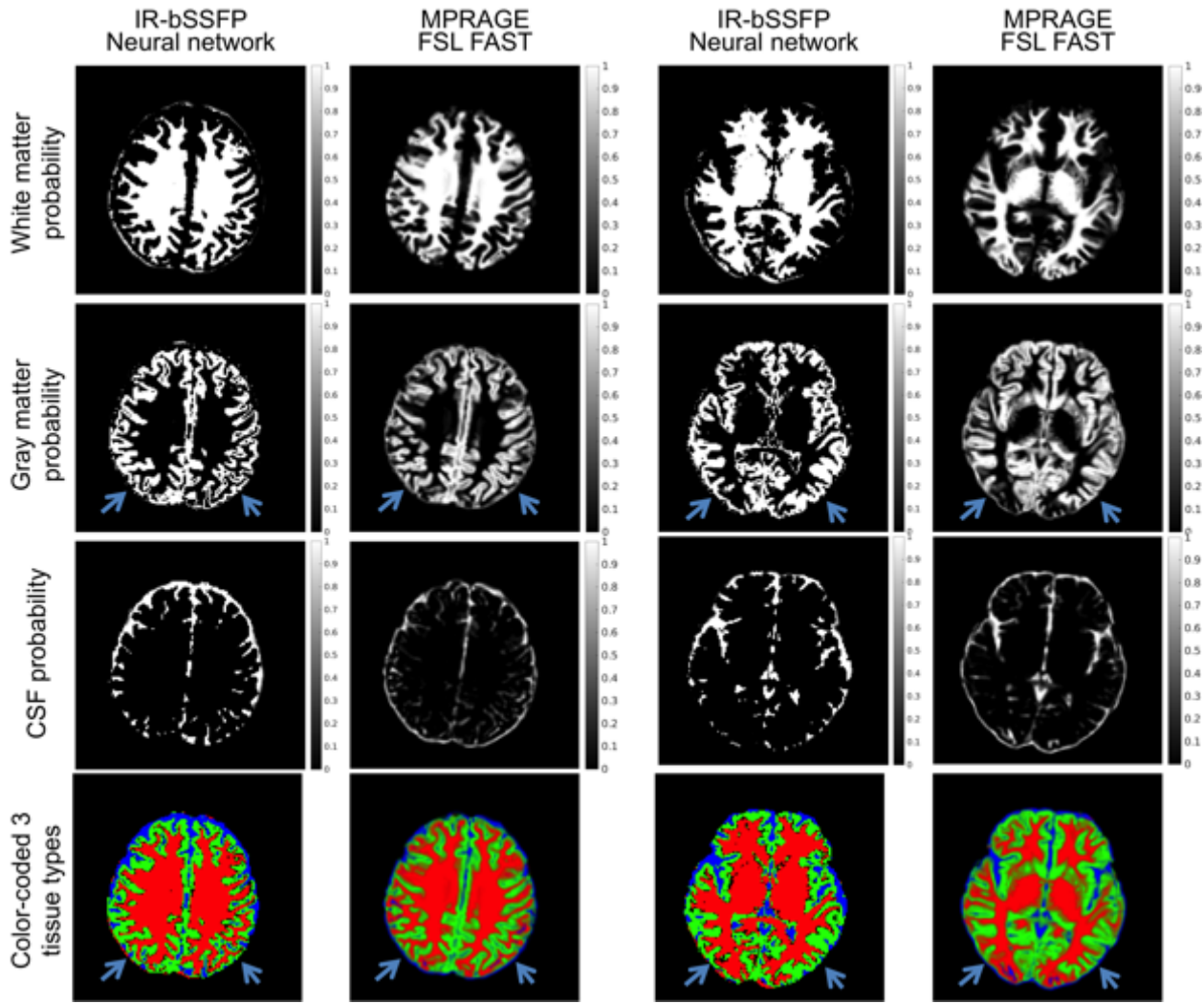

**Figure 6** Whole brain segmentation comparisons between the direct output of the neural network and FSL FAST (same study as Fig. 5). In general, the neural network based segmentation matched FSL FAST

segmentation results. The neural network showed robustness in segmenting gray matter in the cortex (arrows) but did not correctly segment the putamen on this slice. The skull stripping was performed using the BET algorithm in FSL.

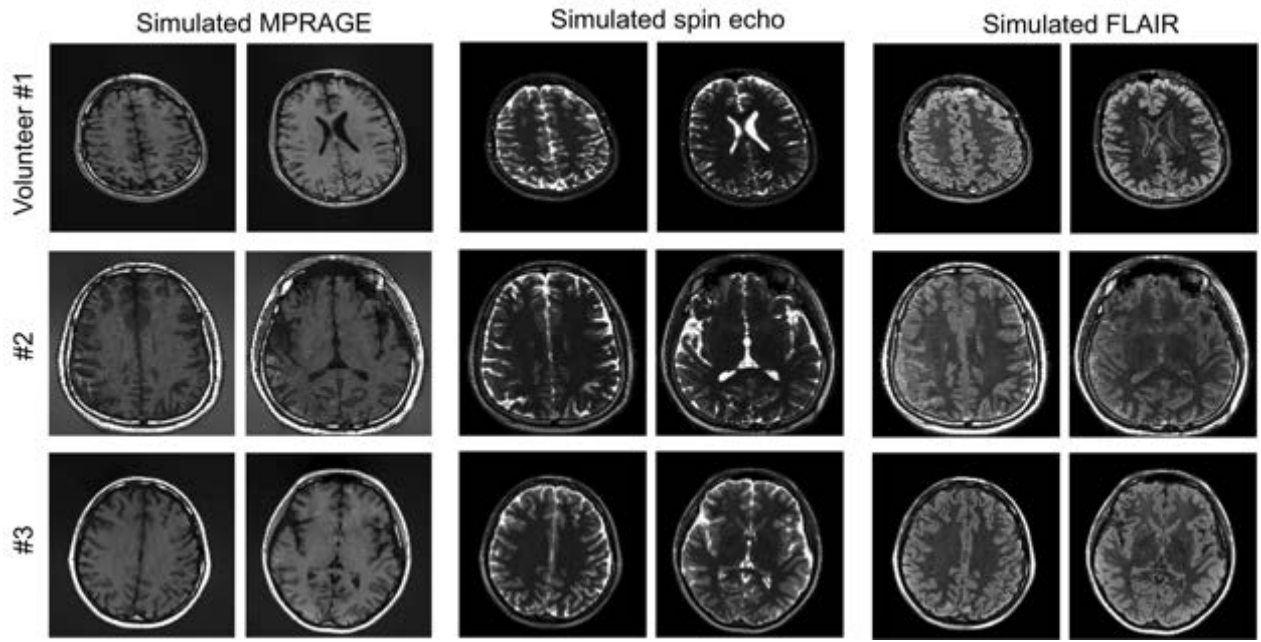

**Figure 7** Simulated MPRAGE images (inversion time = 1500 ms, TR = 3500 ms), spin echo (TE = 80 ms) and FLAIR images (TE/TR = 80/3500 ms, inversion time = 1500 ms), from the T1, T2 and proton density/M0 results of neural network. The bias field correction, with the tissue segmentation input, was applied to the raw proton density/M0 maps before the simulation.

In Figure 8, the potential clinical value of the proposed method was demonstrated in providing significant T2/T1 contrast for ischemic lesions on the rat model. Preliminary results in Figure 8 show the lesion area was automatically segmented by the proposed method due to the abnormally long T1 and T2 in the lesion area. However, the current model cannot differentiate the lesion and CSF because they presented similar T1 and T2 values in the quantification results as shown in the first and second rows (Fig.8).

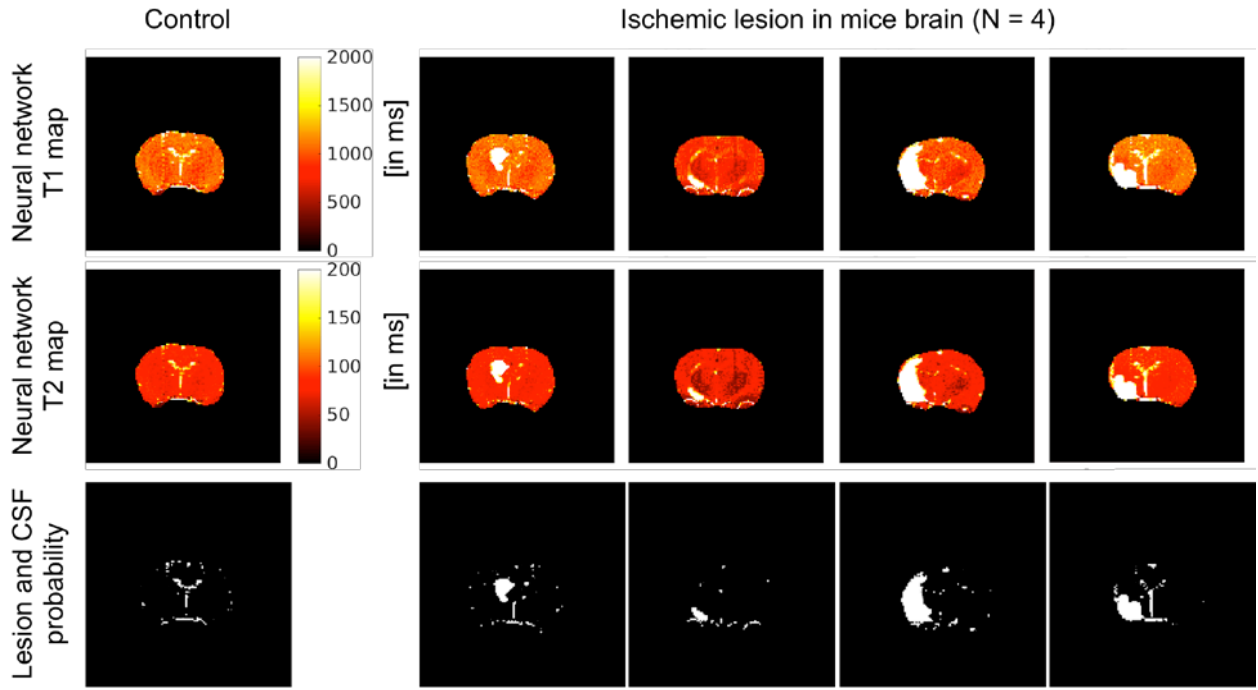

**Figure 8** Animal feasibility study shows the proposed method can provide significant contrast for ischemic lesions on the rat model, allowing semi-segmentation of lesion/CSF area in the neural network probability output.

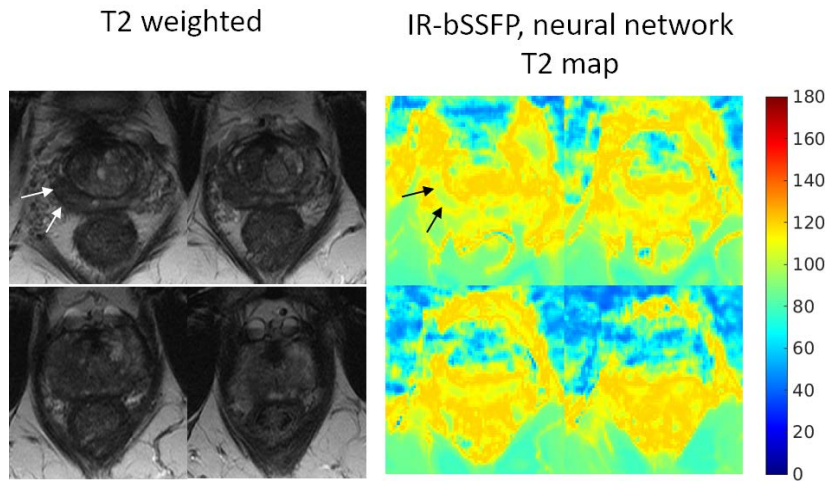

**Figure 9** Preliminary results from IR-bSSFP and neural network shows feasibility of the proposed method for quantitative imaging in the human prostate, where the lesion in the peripheral zone can be detected by the proposed method (arrows). The mosaic plots are from four different slice locations.

**Table 1** neural network estimated apparent T1 and T2 values (in ms, mean ± standard deviation) on three healthy volunteers (V1 to V3) at 3T

|    | Acq details | Relaxations | WM | GM | CSF |
|----|----|----|----|----|----|
| V1 | Without B0 estimation | T1 | 578.6 ± 113.0 | 940.5 ± 314.4 | 2775.2 ± 679.4 |
|    |             | T2 | 38.5 ± 9.4    | 72.3 ± 22.9   | 178.5 ± 61.5   |
| V2 | With B0 estimation | T1 | 601.8 ± 107.6 | 927.6 ± 268.2 | 2290.8 ± 643.9 |
|    |             | T2 | 39.9 ± 8.5    | 76.9 ± 23.5   | 190.6 ± 56.4   |
| V3 | With B0 estimation | T1 | 608.3 ± 117.2 | 983.6 ± 297.2 | 2371.1 ± 722.1 |
|    |             | T2 | 41.5 ± 8.1    | 80.2 ± 23.7   | 167.5 ± 59.0   |
|    | In Ref (15) | T1 | 832           | 1331          |                |
|    |             | T2 | 80            | 110           |                |

**Discussion**

In this study, we proposed that a deep neural network, which learned the MRI signal evolution and tissue-parameter association, could perform a multiple, relatively complicated image analysis tasks of relaxometry, segmentation, and synthetic image generation. To perform the tasks separately, i.e., to quantify the T1 and T2 first and then perform the segmentation, the quantification would cause the scattering artifacts, as shown in Figure 4 when using the naive single-task and dictionary-element like quantization networks. On the other hand, the joint quantification and segmentation demonstrated in this study effectively constrained the T1/T2 values by using limited tissue type value ranges for training (as illustrated in Fig. 1), which helped the algorithm to perform the T1 and T2 estimation. The T1/T2 quantification was improved by taking advantage of the T1 and T2 range limitations. We also found that if without the segmentation task, i.e., applying deep learning only for the T1/T2 quantification, the artifact level was similar to that of dictionary matching (data not shown). An advantage of the proposed multitask learning paradigm is that whole brain relaxometry quantification, and pixel-wise image segmentation can be done within ~5 seconds once the model is trained. This may lead to a major shift from the conventional paradigm, typically "MRI scan->model fitting->T1/T2 parameters->segmentation/analysis", by rapidly and automatically performing both relaxometry and segmentation. Currently, the proposed method or other similar MRF approaches have not been used clinically. Actually, the long quantification time for processing the dynamic MR images in those MRF approaches would limit their future clinical utility. For example, the dictionary matching approach would take 10 to 20 mins to perform the quantification task; therefore, the deep neural network with the capability of tissue quantification and segmentation in 5 seconds could increase the efficiency for data processing workflow, which could encourage the clinical use of this type of approach.

In this initial study, the method was demonstrated on healthy volunteers with fixed T1 and T2 ranges for tissue-parameter associations, but potentially one could include T1 and T2 ranges for pathological tissues. However, such modeling could be challenging for the segmentation of heterogeneous tissues, and requires further development of a more complex neural network with the capacity of spatial structural representation. A potential implementation could be based on a fully convolutional neural network (FCN) [34]. Such neural network may enable auto-assessment of the T1/T2 burden of neurological diseases. Figures 8 and 9 show the application of the proposed method in the semi-automatic segmentation of high T1/T2 lesion areas in the rat model (N = 4) and for relaxometry in human prostate (N=1) in vivo. Those results demonstrated the potential for clinical application and generalizability of the proposed method. In Figure 6, the putamen and the head of caudate were not recognized by the proposed method likely caused by the relatively low spatial resolution that did not allow accurate separation of the mixed grey/white matter voxels. However, with higher spatial resolution, these deep brain structures could be correctly segmented, as shown in the first volunteer study (in Fig. 3).

In this initial study, the magnetization transfer effect is likely to have resulted in the smaller T1 and T2 values compared with those from standard T1 and T2 mapping (in Fig. 5), and literature values [15, 16]. Intuitively, the bond water pool with short T1 and T2 is in an exchange with the free water pool. The magnetization saturation for the bond water from inversion and excitation pulses would be transferred to the free water, causing a pseudo faster relaxation of IR-bSSFP signal. It was known that the longer pulse width, which has a narrower bandwidth, will result in less magnetization transfer effect in bSSFP readout [31–33]. However, to fully characterize the magnetization transfer effect, Bloch-McConnell simulation is needed; this will be investigated in further experiments. In addition, the proposed model has not been tested on human brain data. To summarize, these issues would be solved by further development of more advanced neural networks and more complex synthetic MRI simulations with a similar training procedure and architecture.

**Conclusion**

In conclusion, the proposed deep neural network method trained with MR signal simulations can directly generate apparent T1 and T2 maps as well as synthetic T1 and T2 weighted images in conjunction with segmentation of gray matter, white matter and CSF.

**Acknowledgment**

This work is supported by grants NIH R56HL133663 and NIH R21NS089004.

**Figures captions**

**Figure 1 (a**) MRI signal synthetic model in the brain is used to train the neural network; while in vivo MRI data is used to test the neural network. Specifically, the proposed method assumed the feasible parameter space could be divided into boxes that were binary labeled for white matter (WM), gray matter (GM), and cerebrospinal fluid (CSF). **(b)** A deep neural network was used to perform pixel-wise estimation of T1, T2, B0 (not shown) and proton density/M0 (weighted by coil sensitivity variation) as a regression output, and WM, GM and CSF probabilities as a classification output. Each input pixel contained coefficients of principal components from a principal component analysis (PCA) preprocessing. The neural network contains 3 residual blocks, and each block has 8 fully connected layers, i.e. an MLP with "by-pass" connections to avoid gradient vanishing during training. The neural network was trained with T1 and T2 values from the parameter feasibility region as well as varied proton density weightings, varying noise levels, and binary class labels (in Fig. a).

**Figure 2** The comparison of the proposed method with T1 and T2 mapping by standard parameter mapping methods (i.e., IR-FSE and CPMG) on six Sprague Dawley rats (N = 6 with two normal controls and four with chronic ischemic lesions in the left hemisphere, all images and analysis are presented in Supporting Figure 1). There is a strong spatial agreement between the methods. The differences/errors (mean difference ± standard deviation of difference) in two Bland Altman plots were 113 ± 486 and 154 ± 512 ms for T1, and 5 ± 33 and 7 ± 41 ms for T2, respectively. The on average shorter T1 and T2 values resulting from IR-bSSFP with neural network processing can be attributed to magnetization transfer effects [31–33].

**Figure 3** The segmentation and relaxometry results for one slice (out of 30 slices). The data were acquired with 3D IR-bSSFP, 13 inversion times, and no frequency offsets. Notice that the proposed method tolerated varied coil sensitivity profiles and the modest artifacts in the raw images caused by the aliasing and Gibbs ringing along slice dimension (top left). The estimated T1 and T2 maps are consistent with our knowledge of brain anatomical structure in MRI. The 3D Markov random field algorithm was applied to T1 and T2 maps for a segmentation comparison. The classification result from Markov random field showed some ambiguities in detecting detailed white matter structures (arrow), likely due to the error propagation from T1 and T2 maps. The skull stripping was performed using the BET algorithm in FSL.

**Figure 4** In vivo apparent T1, T2 and B0 estimation using the proposed multi-task deep neural network with T1 and T2 range limitations, single-task deep neural network with T1 and T2 range limitations, and dictionary-element-like quantization on T1, T2, and B0, as well as the naive single-task approach with different levels of T1 and T2 limitations. Data were acquired with 3D IR-bSSFP, 30 inversion times, and 2 different frequency offsets at 0 and 1/(2×TR). The single-task and dictionary-element like quantization were vulnerable to artifacts in T1, T2, and B0 maps. On the other hand, the deep neural network outputs with T1 and T2 range limitations appeared more accurate relative to brain anatomy and provided more continuous T1, T2 and B0 values.

**Figure 5** Comparison of the proposed method with T1 and T2 mapping by standard parameter mapping methods (top row). There is a strong structural agreement between the methods. The on average shorter T1 and T2 values resulting from IR-bSSFP with neural network processing can be attributed to magnetization transfer effects [31–33].

**Figure 6** Whole brain segmentation comparisons between the direct output of the neural network and FSL FAST (same study as Fig. 4). In general, the neural network based segmentation matched FSL FAST segmentation results. The neural network showed robustness in segmenting gray matter in the cortex (arrows) but did not correctly segment the putamen on this slice. The skull stripping was performed using the BET algorithm in FSL.

**Figure 7** Simulated MPRAGE images (inversion time = 1500 ms, TR = 3500 ms), spin echo (TE = 80 ms) and FLAIR images (TE/TR = 80/3500 ms, inversion time = 1500 ms), from the T1, T2 and proton density/M0 results of neural network. The bias field correction, with the tissue segmentation input, was applied to the raw proton density/M0 maps before the simulation.

**Figure 8** Animal feasibility study shows the proposed method can provide significant contrast for ischemic lesions on the rat model, allowing semi-segmentation of lesion/CSF area in the neural network tissue/fluid probability outputs.

**Figure 9** Preliminary results from IR-bSSFP and neural network shows feasibility of the proposed method for quantitative imaging in the prostate, where the lesion in the peripheral zone can be detected by both methods (arrows).

**Supporting Figure 1** The comparison of the proposed method with T1 and T2 mapping by standard parameter mapping methods (i.e., IR-FSE and CPMG). There is a strong structural/spatial agreement between the methods. The differences/errors (mean difference ± standard deviation of difference) in twelve Bland Altman plots were 113 ± 486, 286 ± 342, 154 ± 512, 140 ± 600, 120 ± 607, and 71 ± 611 ms for T1, and 5 ± 33, 6 ± 21, 7 ± 41,2 ± 31, 13 ± 37, and 4 ± 37 ms for T2, respectively. The on average shorter T1 and T2 values resulting from IR-bSSFP with neural network processing can be attributed to magnetization transfer effects [31–33].

**Supporting Figure 2** (**a**) Typical inputs for the bias field correction: proton density/M0 map (with the coil sensitivity variation and aliasing artifacts), white matter, gray matter, and CSF probabilities. (**b**) Effective bias field correction was achieved with the use of brain segmentation.